# Design and Implementation of an Intelligent Educational Model Based on Personality and Learner's Emotion

Somayeh Fatahi

Department of Computer Engineering
Kermanshah University of Technology
Kermanshah, Iran
Email: fatahi_somayeh@yahoo.com

Nasser Ghasem-Aghaee

Department of Computer Engineering
Isfahan University
Isfahan, Iran
Email: nasser_ga@yahoo.ca

*Abstract*—**The Personality and emotions are effective parameters in learning process. Thus, virtual learning environments should pay attention to these parameters. In this paper, a new e-learning model is designed and implemented according to these parameters. The Virtual learning environment that is presented here uses two agents: Virtual Tutor Agent (VTA), and Virtual Classmate Agent (VCA). During the learning process and depending on events happening in the environment, learner's emotions are changed. In this situation, learning style should be revised according to the personality traits as well as the learner's current emotions. VTA selects suitable learning style for the learners based on their personality traits. To improve the learning process, the system uses VCA in some of the learning steps. VCA is an intelligent agent and has its own personality. It is designed so that it can present an attractive and real learning environment in interaction with the learner. To recognize the learner's personality, this system uses MBTI test and to obtain emotion values uses OCC model. Finally, the results of system tested in real environments show that considering the human features in interaction with the learner increases learning quality and satisfies the learner.**

## I. INTRODUCTION

One of the most important applications of computers is virtual learning. In recent years, many organizations have started to use distance learning tools. Although this type of education has some advantages, they don't deal with sufficient dynamism and often the education systems do not have any capability of a real class [29]. Nowadays, an effort is being made to make the virtual learning environments as real-like as possible using intelligent agents with emotions and a personality as well as simulating human behavior.

It's clear that learners' emotional states change during learning processes; the changes in the learning process depend on individual differences and events that happen in the environment [9]. Positive emotions play an important role in creativity and flexibility for solving problems. On the other hand, negative emotions block the thinking process and prevent sound reasoning [7]. For example, if the students are tired and stressful, they cannot concentrate on lessons and think properly [24] [29]. So, the emotional states of learners in virtual learning environments must be taken into account [7]. Besides, people have different personalities, and these differences absolutely affect their duties and daily activities. People with different personalities show different emotions in facing events. Also, different personalities play an important role among learners in the learning process. Personality of learners can affect their learning styles [13]. According to their personalities, each person has especial learning style, and therefore the teaching style that must be used for every student varies from student to student.

In virtual learning systems that have been made until now, scientists have focused on the learner's emotions and have used emotional agents. Personality is an independent parameter in few of them. There is an important problem that has not been solved until now, there is some models in this field, but no one of them have perfect view relative to learner's personality and emotions together.

This paper is organized in the following ways: section 2 is a review of the previous works and literature. Section 3 explains psychological principles. Section 4 is about the proposed model, and section 5 discuses implementation of the model. Finally, section 6 and 7 explain evaluation of the system, results and future works.

## II. PREVIOUS WORKS

Many researchers had tried to design general models for emotions in artificial intelligence area [28] [36] [41] [16] [34]. Due to the fact that emotions' effect on the learner plays an inevitable part in the real world, if we neglect them, it's counted as a big fault in virtual learning systems [2].

There are many efforts in modeling of emotions in the field of virtual learning. Some of these efforts are explained here: Barry Kort, Rob Reilly, Rasalind Picard have presented a model whose target is using the effect of emotions on learning; then, they implemented their models so that their system could recognize emotional state of learners and respond to them according to the state [34]. Shaikh Mostafa Al Masum and Mitsuru Ishizuka have presented an emotional cubic model. The cubic model has three dimensions. According to these three dimensions, eight types of emotions are extracted that are effective in the learning process. In this model, they have used fuzzy science for emotional states [2]. Qiao Xiangjie and his colleague have presented a self-assessment model. In this model, the learner finds his/her emotional state by data that are gathered from the self-assessment based on an emotion map. They have used a polar model for extracting emotions [46].

Some people just have presented personality models, such as Ushida and his colleague who have modeled personality types based on differences in individual emotional states. Rosis and his colleague have modeled and implemented the personalities according to the change of the agents' priority of goals .Serra and Chittaro have suggested a goal bound method for modeling agents. Ball and Breese have modeled intensity of emotions in the agent with two personality features in a BBN (Bayesian Belief Network) network. Thalmann and Kishirsagar have used BBN for modeling features. They have used FFM personality model and OCC emotional model, and have added surprise and hate emotions to it [44]. Andre and his colleagues have presented an integrated model of emotions (based on OCC model) and personality (based on FFM model). At the beginning, they have simulated basic emotions like sadness, joy, fear and anger, and two personality dimensions: extroversion and pleasantness [44] [30]. Jin Due and his colleagues tried to obtain modeling from the learner. In their model, the learner's personality is extracted based on Cattell questionnaire, and then the relation between personality and behavior is obtained using data mining techniques [13].

After reviewing emotion and personality models in the virtual learning, we will explain some systems that use this model: Chaffer, Cepeda and Frasson have tried to predict the learner's emotion in E-learning systems. They have used Naïve Bayes Classifier method for predicting and modeling the learner's emotional reaction according to his/her personal features such as sex, personality, etc. [7] [8]. Ju and his colleagues have designed a software agent that can cooperate with the learner. They put this agent in a non-synchronic learning environment. The designed system included two subsystems: the teacher subsystem and the learner subsystem. In the learner subsystem, learners are grouped according to their level of knowledge and willingness to cooperate, and are treated accordingly. Keleş and his colleagues have presented a learning system that is called ZOSMAT. The system is implemented in a real class environment. The system is a learner based system [44]. Chaffer and Frasson have suggested ESTEL architecture to determine optimal emotional state of learning and induct it to the leaner. The architecture uses Naïve Bayes Classifier method to predict the optimal emotional state of the learner based on his/her personality. The optimal emotional state has been defined so that it increases the learner's efficiency, then the optimal emotional state is inducted to the learner using some combined techniques such as pictures, music, etc. [24]. Chalfoun, Chaffar and Frasson have designed an agent that can predict the emotional state of the learner in the e-learning environment. The prediction of the agent based on personal and impersonal features is done through a machine learning technique called ID3. Also they used OCC emotional model [9]. Haron have suggested a learning system with a learning module, which can adapt to each learner. This environment uses fuzzy logic and MBTI personality test [1]. Passenger software is designed by Marin and his colleagues to be used for laboratory lessons in distance education. OCC model is used for implementation of the software. The system uses virtual tutor agent [35]. Ju and his colleagues have implemented the learning environment, and have studied the effects of efficiency of virtual classmate agent on the learner's efficiency. In their environment, the virtual classmate agent has either a competitor or a cooperator [25]. Maldonado and his colleagues have designed a system that provides an environment to provide learning through interaction with software agents. The software agent, who acts as a classmate in the system, tries to answer the learner emotionally. This system is used a cooperated agent to help learner [35].

## III. PSYCHOLOGICAL PRINCIPELS

Emotion, personality and individual differences are effective parameters in human activities especially in learning. Every person has special learning style according to his/her personality features [3].

### A. Emotion

Emotions are our reactions to the surrounding world. Aristotle, defined emotion as "that which leads one's condition to become so transformed that his judgment is affected and, which is accompanied by pleasure and pain"[21]. Damasio have proven that the emotions affect reasoning, memorizing, learning and decision making [11]. Studying has showed that intelligence is effective in learning process as much as emotion, interest rate and individuals do [29]. Other people such as Bower and Cohen believe that emotions affect remembering and decision making [8] [29].

#### 1) OCC Model

A lot of models have been designed for emotions. They help us implement emotions and peoples' reactions in different conditions such as encountering an event. One of the most famous is OCC model that is used in most researches. This calculating model is established by Ortoney, Clore and Collins in 1998. The model determines 22 types of emotions. Emotions are divided into positive and negative ones, based on positive or negative reactions to events. The OCC model is calculated intensity of emotions based on a set of variables. The variables are divided into two groups: global and local. Global variables affect all the emotions, however; local variables affect just some emotions. Global variables include senses of reality, proximity, unexpectedness and aroused local variables include desirability, praise worthiness and attraction. The other local variables include desirability for others, deservingness, liking, likelihood, effort, realization, strength of cognitive unit, expectation deviation and familiarity [28].

The OCC model has three branches. The first branch is the emotions which show the result of happening events. These results are obtained according to desirability or undesirability level of events compared to the agents' goals. This branch includes four classes and twelve emotions (Happy-for, Resentment, Gloating, Pity, Hope, Fear, Joy, Distress, Satisfaction, Disappointment, Relief, and Fear-confirmed) [4].

The second branch is emotions that are pointed out the result of agent function based on approving or disapproving relative to a set of standards. The second branch has just one class. It includes four emotions (Pride, Shame, Reproach, and Admiration).

The third branch consists of emotions that are the consequence of the agent's liking or disliking his/her goals compared to the agent's position and attitude. This branch has just one class that includes two emotions (Love and Hate).

There is still another class beside these three branches, and includes four compound emotions (Anger, Gratitude, Remorse, Gratification) [38].

### B. Personality

There are various psychological definitions for personality. In Schultz's view, the unique, relatively constant internal and external aspects of a person's character that influence his behavior in different situations are called personality. In fact, personality includes thoughts, feelings, wishes, inclinations, and behavioral tendencies that are ingrained in various aspects of each person's existence [21].

#### 1) Learning Style

The psychological studies show that each person displays several individual features in problem-solving and decisions-making. These features are often considered as learning styles or learning methods [33]. It's clear that each person has a specific learning style according to his/her personality features [47]. According to the Keefe's definition of learning styles [32], "the learning style is composed of cognitive and emotional characteristics and factors of every individual and is applicable as a set of permanent indices for recognizing how the learner comprehends the concepts, and interacts with the learning environment and responses to the environment". Learning styles are the criteria in perception of information and evaluation of understanding them [1] [14] [32] [47]. There is an important point in the definition, that is, the learning styles reflect preferences and individual priorities in selection of learning conditions [12] [14].

#### a) Evaluation of Learning Styles

Different tools are used to determine learners' learning styles [3]. There are many questionnaires that categorize each person according to their learning styles: Kolb questionnaire, Honey and Mumford questionnaire [33], GRSLSS questionnaire [33] [31], etc.

#### b) MBTI (Myers-Briggs Type Indicator)

In comparison with the questionnaires in other education sciences, MBTI is known as a strong instrument to determine learning styles of individuals [14]. It is an evaluation instrument related to Jung personality theory, the first time, it used by Kathrin Briggs and Isabel Myers Briggs in 1920 [42] [14]. It was first used as a job applicants' evaluation test. After that it was used in education sciences in 1957[43]. Generally this questionnaire is used as an instrument for education and business to determine learning or teaching styles, communicating styles and job selection [42] [12]. According to MBTI grouping, every person has instinctive priorities that are decisive in their behaviors in different conditions [12] [14]. The questionnaire helps to specify the personality features and learning priorities of each person, and to extract the teaching styles are related to the features [40]. MBTI uses four two-dimensional functions according to the Jung theory. Jung theory specifies three functions of Extraversion/Introversion (E/I), Sensing/Intuition (S/N) and Thinking/Feeling (T/F), but we have a fourth dimension, that is, Judging/Perceiving (J/P) in MBTI [1] [40].

#### c) The Features of "MBTI" Dimensions

*Extroversion/Introversion:* extrovert people tend more to the outside world. In fact, they have many tendencies for teamwork. They have a lot of friends. They are active and practical. Their emotions are easily expressed [12] [14] [22] [40]. Conversely, introvert people prefer their introverted opinions and internal world and ideas. They are very independent. They spend a lot of time to think on their tasks; they have a few friends. They try to hold their emotions and express their emotions at certain times to particular people [14] [22] [40] [45].

*Sensing/Intuition (S/N):* sensing people are emotional people who get information from environment through their five senses. They are realists. They usually pay attention to the details, focus on practical subjects [14] [22] [24] [45]. On the other hand, intuitive people, who get information through perception between relationships and results, usually use their conception to get information. They try to make a mental picture of the subject for themselves and then move towards details. Their concentration is more on ideas and their integrity. Their concentration is on the future rather than present [14] [22] [24] [45].

*Thinking/Feeling:* thinking people make their decisions based on exact data, and they like accurate subjects. Their decisions are logical and impersonal [33] [22] [24] [45]. Feeling people, on the other hand, have emphasis on harmony and balance. They enjoy teamwork. Their judgments and decisions are based on personal value.

*Judging/Perceiving:* judging people prefer completely organized life and regulated thoughts and ideas. They pay attention to activities which are important to them. Deadlines are important for them [14] [22] [45]. Perceiving people, however, have a flexible life style. They are curious, agreeable and tolerant. They start several projects simultaneously. They don't pay attention to deadlines.

#### d) Personality Types of "MBTI"

Sixteen personality types result from mixing four two-dimensional functions and individuals are categorized in these types after completing some questionnaires [12] [24]. The sixteen groups are shown in Table I. For example, people in ENTP group are all extrovert, intuitive, thinking and perceiving.

TABLE I.    PERSONALITY TYPES OF "MBTI"

| ESTJ | ISTP | ENTJ | INTP |
|------|------|------|------|
| ISTJ | ESTP | INTJ | ENTP |
| ISFJ | ESFP | INFJ | ENFP |
| ESFJ | ISFP | ENFJ | INFP |

## IV. PROPOSED MODEL

In this paper, a new model is presented according to the learning model based on emotions and personality [19] and virtual classmate model [17] [18] in our previous research. This module is displayed in Figure1.

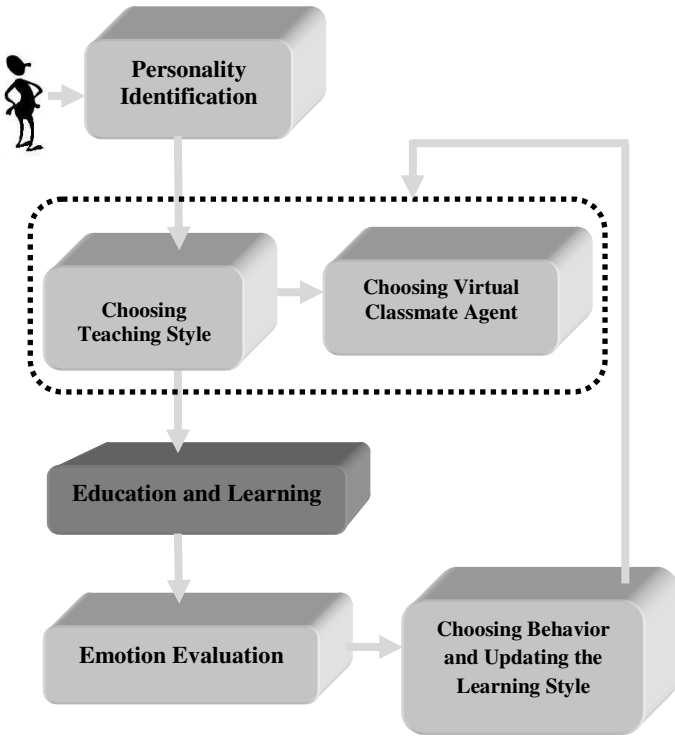

Figure 1.    Proposed model

The model has goals as following:

- Presentation method to determine learner's group based on personality questionnaire

- Presentation method to recognize learner's emotions based on his/her personality

- Presentation method to change teaching style based on learner's emotions and personality

- Presentation a method to determine VCA based on learner's personality

- Presentation suitable tactics in learning based on learner's emotions and personality using VTA and VCA.

### A. Model Architecture

The model includes six modules, each of them described below:

*Module of personality identification:* In the first, learner encounters with system, MBTI questionnaire is in front of him/her, and it finds out learner's personality (for example ISEJ, ESTP, INTJ …).

*Module of choosing a teaching style matching with learner's personality:* Generally, there are three kinds of learning environment: Independent, cooperative, and competitive learning environment [15]. The system sets learner in one of three kinds of independent group, cooperative group with VCA, or competitive group with VCA. Individuals are classified according to personality types that are showed in Table II. This classification is obtained based on fulfilled studies on personality's characteristics [23].

*Module of choosing virtual classmate agent:* If the learner be placed in the independent group, the education and learning process will be started otherwise system selects a suitable VCA according to learner's personality type, then; the process of education and learning will be started. Selecting personality of VCA should be in such a way that is proportionate with the leaner's personality type and it helps to progress of learner during the learning process. According to the fulfilled studies, presence of VCA with an opposed personality is suitable [6] [10] [39]. Studies in [6] [10] [39] and other psychological sources are showed that people are grouped with opposed personality are much better than the people are grouped with similar personality. Former has high efficiency rather than second.

TABLE II.    LEARNER'S CLASSIFICATION BASED ON MBTI TYPE

| Independent | Cooperative with VCA | Competitive with VCA |
|-------------|----------------------|----------------------|
| INTJ | INFJ | ENTP |
| INFP | ENFP | ENTJ |
| ISTJ | ISFJ | ESFJ |
| INTP | ESFP | |
| ISFP | ESTP | |
| ISTP | ENFJ | |
| | ESTJ | |

*Module of education and learning:* In this module, lesson's points are presented to learner as exercises and learning process will be started.

*Module of emotion evaluation:* During doing exercise and evaluating the rate of learning, based on the learner's learning level and the events which happen in the environment, some emotions are expressed in the learner (e.g. liking VCA, disappointment from doing exercises, etc). According to the fulfilled studies, we have found that only specific emotions are effective in the learning process. The result of the studies in [29] [36] are showed that the first branch of emotions in the OCC model is effective in learning. Here, we use first and third branch of OCC model. The first branch includes effective emotions in the learning process, and the third branch includes those emotions that the person in the relation with the others (e.g. the virtual classmate agent) shows.

_Module of choosing behavior and updating style of teaching:_ According to the events which happened in the environment and was changed learner's emotions; this module based on personality's characteristic and current learner's emotions, changes the teaching style. In addition, VTA and VCA perform suitable tactics to interacting with learner based on event has happened. The tactics are in the knowledge base of system and include a set of rules like "if-then".

### B. Calculation of Emotions' Values

In the proposed model, two emotion's branches of the OCC model are used. The first branch includes effective emotions in the learning process and third branch includes person's emotion against the others. In this part, we examine the way of calculating them.

#### 1) First Branch of Emotions in OCC Model

We use calculating model to achieve emotions' values [26]. In this model to calculate the first branch's emotions of OCC model, values of general variables (Unexpectedness), and local variables (Likelihood and Desirability) should be calculated. Unexpectedness and Likelihood variables directly are obtained from outside, and calculate based on environmental variables. In implementation, which is based on this model [27] as a VTA, the value of Unexpectedness and Likelihood variables have been calculated according to the learner's answer for the exercises (e.g. True or False) [26]. Here, in the same approach, calculating the values of these two variables is done, but calculation value of Desirability is different.

##### a) Learner Agent's Goals According to Personality Type

According to the model [26], when an event happens in the environment, this event is offered to event's evaluating module. According to the learner's goals, evaluation is done and the value of Desirability is obtained.

In this paper, according to the "MBTI" questionnaire, four goals for each learner are considered:

- The goal of learner's agent is to do the exercise alone and without any help from VCA.

- The goal of learner's agent is effort in doing the exercise, even if desirable result isn't obtained.

- The goal of learner's agent is having high speed in answering for the exercises.

- The goal of learner's agent is cooperating with VCA for doing the exercises.

The values of each goal are obtained according to several questions in MBTI questionnaire. After normalization, these values are placed in the numerical domain of 0-1 Eq. (1). These values in the matrix G which is the goals matrix are placed as following [26]:

$$G=\begin{bmatrix} g_1 \\ g_2 \\ g_3 \\ g_4 \end{bmatrix} \quad \forall i \in [1,4]: g_i \in [0,1] \quad (1)$$

In Eq. (1), $g_i$ is valued with goal importance.

##### b) Event's Impact on Learner's Goals

In this stage, the value of each event's impact on the goal is expressed, by a matrix Eq. (2):

$$\text{Impact}=\begin{bmatrix} a_{11} & ... & a_{1n} \\ & . & \\ & . & \\ a_m & ... & a_{mn} \end{bmatrix} \forall i \in [1,n], \forall j \in [1,m]: a_{ij} \in [-1,1] \quad (2)$$

Each element of the matrix shows value of each event's impact on a special goal. "m" includes number of events and "n" includes number of goals. Positive values of $a_{ij}$s show the positive effect of event on the goal and negative value of $a_{ij}$s shows the negative effect of event on the goal. The values of each $a_{ij}$s are calculated for $i_{th}$ event and $j_{th}$ goal. To calculate each of $a_{ij}$s, the "Value Function' is used [26].

To calculate the "Value Function" environmental variables are used. These variables are used to measure the achievement to goal for an event. Two vectors are defined: vector "V" which shows the effect of environmental variables' values on a goal, and vector "W" which shows the effect of the weights' values of each environmental variable on a goal. If "n" variables affect a goal, they showed in vector "V", as following Eq. (3):

$$V=\begin{bmatrix} V_1 \\ V_2 \\ . \\ . \\ . \\ V_n \end{bmatrix} \quad \forall i \in [1,n]: v_i \in [0,1] \quad (3)$$

The weight of each variable is also placed in vector "W", as following Eq. (4):

$$W= [w_1 \ w_2 \ ... \ w_m] \quad \forall j \in [1,m]: w_j \in [0,n] \quad (4)$$

Value of the "Value Function" is obtained by multiplying two vectors of "V" and "W". Because of the value of the "Value Function" should be between 0 and 1, we divide obtained product to the sums of the elements of weights' vector Eq. (5):

$$\text{Value } (g_j) = \frac{W \times V}{\sum_{k=1}^{m} W_k} \quad (5)$$

##### c) Environmental Variables

In the proposed model, according to the defined goals for a learner, five environmental variables are considered, these variables influence the goals:

- Independence
- Potential of Cooperation
- Response Speed
- Grade of Exercises
- Effort

According to the fulfilled studies on MBTI questionnaire, for each of the influencing variables on learner's goals, it is considered a numerical weight in the domain of 1-3 [42] [12] [14] [22] [24] [40] [45]. Variables and weights are shown in Table III.

TABLE III.    ENVIRONMENTAL VARIABLES

| *Vector V* | *Vector W* | | | |
|---|---|---|---|---|
| | Goal 1 | Goal 2 | Goal 3 | Goal 4 |
| Independence | 3 | 1 | 1 | 1 |
| Potential of Cooperation | 1 | 1 | 1 | 3 |
| Response Speed | 1 | 1 | 3 | 1 |
| Grade of Exercises | 2 | 1 | 1 | 2 |
| Effort | 1 | 3 | 1 | 1 |

*Independence:* Value of the variable shows the learner's independence during the learning process. At the first learner encounter the system, value of this variable for the cooperative and competitive group is 0 and for the independence group is 1. During learning process, the value of this variable calculated according to Eq. (6):

$$\text{Independence} = 1 - AH \qquad (6)$$

In this equation, "Independence" variable shows learner's independence, and "AH" is indicator of VTA's help or VCA's help. The AH is between 0-1, and it's obtained by the division of aid's request value from VTA or VCA on the number of exercises.

*Potential of Cooperation:* Value of the variable shows learner interesting in cooperative group, the value of this variable calculated according to Eq. (7):

$$\text{Potential of Collaboration} = 1 - \text{Independency} \qquad (7)$$

In this equation, the independence variable is calculated from Eq. (6).

*Response Speed:* Value of the variable shows learner's response speed that is obtained with Eq. (8).

$$\text{Response Speed} = 1 - \frac{RT}{DT} \qquad (8)$$

In equation Eq. (8), "RT" is leaner's response time, and "DT" is the default time of the system for responding to the exercise.

*Grade of Exercises:* Value of the variable is between 0-1, and it shows the learner's grade from solving the exercises.

*Effort:* Value of the variable is obtained with ask of learner. Value of this variable is between 0-1.

### d) Calculating Desirability

After calculation of the goals' values and the impact value of the event on the learner's goals, Desirability value is obtained by multiplying two matrixes: "Impact" and "G" according to Eq. (9) [26]. In this equation $e_i$ represents the entrance event.

Since the numerical results should be between -1 and 1, the obtained result is divided to the sums of the vector "G" elements.

$$\text{Desirability } (e_i) = \frac{\sum_{j=1}^{n} a_{ij}\, g_j}{\sum_{i=1}^{n} g_i} \qquad (9)$$

Finally, we obtain value of the first emotions branch of OCC model from Desirability, Unexpectedness and Likelihood variables. It should be mentioned that the value of Likelihood variable is obtained with attention to the student's performance in the past, and the value of Unexpectedness variable is obtained according to the student's effectiveness.

### 2)    Third Branch of Emotions in OCC Model

For determining Love and Hate emotions, we attend the event's type and current emotions. When an event happens in the environment, either positive or negative emotions appear in the leaner based on this event. List of the events as well as their classification to positive and negative groups are shown in Table IV. In addition, the emotions' classification based on positive or negative are shown in Table V.

TABLE IV.    CLASSIFICATION OF THE EVENTS

| *Positive* | *Negative* |
|---|---|
| Accurate response to the exercise | Inaccurate response to the exercise |
| Student's effort for solving the exercise | Finishing time of responding and Unsolved exercises |
| Student's thinking | Leaving the class |
| Requesting help from VCA | Rejecting help or refusing to cooperate with VCA |
| | Request for the next exercise and leaving the previous exercise unsolved |

TABLE V.    CLASSIFICATION OF THE EMOTIONS IN OCC MODEL

| *Positive* | *Negative* |
|---|---|
| Joy | Distress |
| Hope | Fear |
| Satisfaction | Disappointment |
| Relief | Fear-confirmed |
| Gloating | Pity |
| Happy-for | Resentment |
| Pride | Shame |
| Admiration | Reproach |
| Love | Hate |
| Gratitude | Anger |
| Gratification | Remorse |

In this section, based on the conducted studies in [37], positive and negative emotions, and positive and negative events, we consider four states as follows:

- If the learner's current emotion is negative and a negative event has happened, Hate emotion toward VCA will be expressed.

- If the learner's current emotion is negative and a positive event has happened, Hate emotion toward VCA will be expressed.

- If the learner's current emotion is positive and a negative event has happened, Love emotion toward VCA will be expressed.

- If the learner's current emotion is positive and a positive event has happened, Love emotion toward VCA will be expressed.

According to these states, Love or Hate emotions toward a VCA is obtained and since the numerical value of these two emotions is not important for the rules of our expert system, it is just enough for the it to differentiate them.

## V.    IMPLEMENTATION

We have implemented our model in educational environments. The educational domain in this environment is Learning English Language. For better evaluating of the proposed model, this environment is compared with two other environments. Three environments to examine are as follows:

- Educational environment without emotions

- Educational environment with emotional VTA

- Educational environment with emotional VTA and VCA who have emotions and personality

### A.    Educational Environment without Emotions

This environment is a simple virtual education environment. The learner enters the environment and just tries to solve many exercises. After responding to the questions of a level, the learner is promoted to an upper level. During the learning process, the environment does not provide any guides for the learners, and just the learner's grade is shown on the screen.

### B.    Educational Environment with Emotional VTA

In this environment, VTA tries to get the learner's emotions by simulation of the learner and select the suitable tactics and perform them, using the learner's emotions and the events happening in the environment. VTA also can show emotional states and interact with the learner. Considering the events happening in the environment, the VTA infers the learner's emotions and, based on them, uses suitable tactics for teaching the learner more effectively. The tactics are saved as a set of rules in the VTA's knowledge base. One of them for example is as follows:

| Rule 1: | IF | | Disappointment | IS | High |
|---|---|---|---|---|---|
| | | OR | Disappointment | IS | Medium |
| | | AND | Event | IS | Wrong Answer |
| | THEN | | Teacher-Tactic 1 | IS | Increase-Student-Self-Ability |
| | | AND | Teacher-Tactic 2 | IS | Increase-Student-Effort |

### C.    Educational Environment with Emotional VTA and VCA with Emotions and Personality

This environment is implemented based on the proposed model. The environment uses VTA and VCA together. Two agents based on recognizing learner's emotions, select suitable tactics to interaction with him/her (Figure 2).

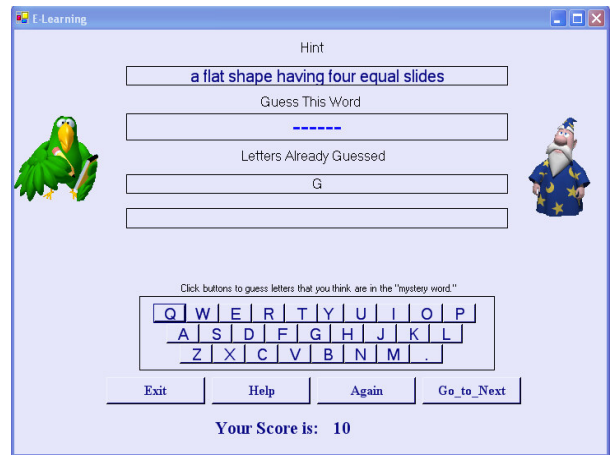

Figure 2.    Educational Environment with VTA and VCA

In this environment, selection of tactics for dealing with the learner is related to the group's type. According to the learner's learning group, certain tactics are used. At the beginning, based on the learner's personality, one of the independent, cooperative, or competitive learning groups is considered for him/her.

The rules of determining the learner's group are saved in the system's knowledge base. For example, one of the rules is as follows:

| Rule 1: | | | | |
|---|---|---|---|---|
| | IF | Student's Personality | IS | ISFJ |
| | THEN | His/her group | IS | Cooperative |

### 1) Independency's Learning Group

When a learner is in this group, he is supposed to solve the exercises by himself and without the presence of VCA. In this environment, VTA doesn't help the learner either and only does his administrative tasks. This learning environment is very similar to the educational environment with the emotional VTA. There is, however, a small difference, that is, since the personality's type of the learner is considered as independent and these people like to solve the exercises by themselves, help with tasks is deleted from the VTA's tasks. The other VTA's tasks are similar to those of the VTA in the educational system as mentioned above. The tactics in the system has been defined for him. The list of VTA's tactics is shown in Table VI.

TABLE VI.     VTA'S TACTICS IN INDEPENDENCY'S LEARNING ENVIRONMENT

| | VTA's Tactics | | |
|---|---|---|---|
| 1 | Increase-Student-Self-ability | 7 | Allow-to-Leave-Virtual-Class |
| 2 | Increase-Student-Effort | 8 | Teacher-IS-Idle |
| 3 | Congratulate-Student | 9 | Propose-Cooperate-With-VCA |
| 4 | Encourage-Student | 10 | Change-Student-Group-to-Cooperative |
| 5 | Recognize-Student-Effort | 11 | Change-Student-Group-to-Competitive |
| 6 | Show-Student-New-Skills | 12 | Show-Next-Exercise |

VTA performs some tactics to interact with the learner. The tactics are saved as rules in VTA's knowledge base. For example, one of the rules is as follows:

| Rule 1: | IF | Student Group | IS | Independent |
|---|---|---|---|---|
| | AND | Disappointment | IS | High |
| | OR | Disappointment | IS | Medium |
| | AND | Event | IS | Wrong Answer |
| | THEN | Teacher-Tactic 1 | IS | Increase-Student-Self-Ability |
| | AND | Teacher-Tactic 2 | IS | Increase-Student-Effort |
| | AND | Teacher-Tactic 3 | IS | Change-Student-Group-to-Cooperative |

#### a) Physical and Verbal Behaviors

Every tactic includes a set of physical and verbal behaviors. The physical behaviors are those emotional behaviors that the VTA expresses for showing his emotional states. These behaviors are executable by the prepared functions of Merlin agent which Microsoft Company has designed.

For each verbal behavior, there is a sentence that VTA presents as text. Instances of the physical and verbal behaviors are shown in Table VII and Table VIII, respectively.

TABLE VII.     PHYSICAL BEHAVIORS OF VTA IN INDEPENDENCY LEARNING ENVIRONMENT

| Physical Behavior | Merlin Function |
|---|---|
| Congratulation | Congratulate, pleased, Speak |

TABLE VIII.     VERBAL BEHAVIORS OF VTA IN INDEPENDENCY LEARNING ENVIRONMENT

| Verbal Behavior | Merlin's Speaking |
|---|---|
| Congratulation | Uauuuuu! You are very well! Congratulations for the efforts that you made! |
| Congratulation | Congratulations! You obtain an excellent result! Continue it. |
| Congratulation | Congratulations! Your performance was stupendous! |
| Congratulation | Congratulations for your efforts! |
| Congratulation | Congratulations! You reached a good result! |

### 2) Cooperative Learning Group

When a learner is in this group, according to his personality's type, he is interested in participating in the group and doing cooperative activities. Considering this issue, the environment has been designed in such a way that it uses a VCA to interaction with the learner.

In this environment due to the presence of VCA, VTA's tasks are changed, and we have defined a set of tasks for VTA and VCA. Considering this fact, the environment is defined so that a VCA is used to communicate with the learner. The tactics of VTA and VCA in Cooperative Learning Environment are shown in Table IX.

TABLE IX.     TACTICS OF VTA AND VCA IN COOPERATIVE LEARNING ENVIRONMENT

| VCA's Tactics | VTA's Tactics |
|---|---|
| Increase-Student-Self-ability | Congratulate-Student |
| Increase-Student-Effort | Recognize-Student-Effort |
| Persuade-Student-To-Think-More-For-Problem | Show-Student-New-Skills |
| Cooperate-With-Student | Show-Next-Exercise |
| Notify-Student-For-Deadline | Allow-To-Leave-Virtual-Class |
| Encourage-Student | Teacher-IS-Idle |
| Give-Help | Change-Student-Group-To-Independent |
| Persuade-Student-To-Be-Independent | Change-Student-Group-To-Competitive |
| Offer-Cooperation | |

According to the above description, a set of rules are saved in the knowledge base of the environment so that the VTA and VCA have interaction with the learner in accordance with them. For example, one of the rules is as follows:

| Rule 1: | IF | | Student Group | IS | Cooperative |
|---|---|---|---|---|---|
| | | AND | Like | IS | High |
| | | OR | Like | IS | Medium |
| | | AND | Distress | IS | High |
| | | AND | Event | IS | Wrong-Answer |
| | | AND | Student's Response speed | IS | Higher-Than-Threshold |
| | | AND | Virtual Classmate's Personality | IS | IN |
| | | OR | Virtual Classmate's Personality | IS | IS |
| | THEN | | Teacher_Tactic1 | IS | Recognize-Student-Effort |
| | | AND | Classmate_Tactic1 | IS | Persuade-Student-to-Think-More-for-Problem |

In this part, VTA's behaviors are executable by Merlin agent, and VCA's behaviors are executable by Peedy agent. Instances of the physical and verbal behaviors, which are related to the VCA, are given in Table X and Table XI.

TABLE X.  PHYSICAL BEHAVIORS OF VCA IN COOPERATIVE LEARNING ENVIRONMENT

| Physical Behavior | Peedy Function |
|---|---|
| Persuade_Student | Think, Speak |

TABLE XI.  VERBAL BEHAVIORS OF VCA IN COOPERATIVE LEARNING ENVIRONMENT

| Verbal Behavior | Peedy's Speaking |
|---|---|
| Persuade-Student | You have to think more for doing your assignment. |
| Persuade-Student | I know, you are very intelligent, but you'd better spend more time on your assignments. |
| Persuade-Student | Let think more on this idea, please. There are enough times. Don't festinate to answer. |

### 3)  Competitive Learning Group

When a learner is in this group he/she may be interested in competing with his classmate, based on his personality type. Concerning this issue, the environment has been designed in such a way that it uses a VCA for competing with the learner and simulates a competitive environment for the learner.

In this environment, due to the presence of VCA, VTA's tasks are changed, and we have defined a set of tasks for VTA and VCA. Considering this fact, the environment is defined so that a VCA is used to communicate with the learner. The tactics of VTA and VCA in Cooperative Learning Environment are shown in Table XII.

TABLE XII.  TACTICS OF VTA AND VCA IN COMPETITIVE LEARNING ENVIRONMENT

| VCA's Tactics | VTA's Tactics |
|---|---|
| Increase-Student-Effort | Congratulate-Student |
| Persuade-Student-To-Think-More-For- Problem | Congratulate-Classmate |
| Notify-Student-For-Deadline | Increase-Student-Self-Ability |
| | Encourage-Student |
| | Recognize-Student-Effort |
| | Show-Student-New-Skills |
| | Show-Next-Exercise |
| | Allow-To-Leave-Virtual-Class |
| | Teacher-IS-Idle |
| | Change-Student-Group-To-Independent |
| | Change-Student-Group-To-Cooperative |

According to the above description, a set of rules are saved in the knowledge base of the environment so that the VTA and VCA have interaction with the learner in accordance with them. For example, one of the rules is as follows:

| Rule 1: | IF | | Student Group | IS | Competitive |
|---|---|---|---|---|---|
| | | AND | Like | IS | High |
| | | OR | Like | IS | Medium |
| | | AND | Distress | IS | High |
| | | AND | Event | IS | Wrong Answer |
| | | AND | Student's Response speed | IS | Higher–Than–hreshold |
| | | AND | Virtual Classmate's Personality | IS | IN |
| | | OR | Virtual Classmate's Personality | IS | IS |
| | THEN | | Classmate_Tactic1 | IS | Persuade-Student-to-Think-More-for-Problem |
| | | AND | Teacher_Tactic1 | IS | Recognize-Student-Effort |
| | | AND | Teacher_Tactic2 | IS | Change-Student-Group-to-Cooperative |

In this part, like the previous part, for each tactic of the VTA and VCA, a set of physical and verbal behaviors are defined. For a better comparison between the two cooperative and competitive states of the virtual classmate agent, it is given the tactics of persuade-student-to-think in Tables XIII and XIV, respectively.

TABLE XIII.  PHYSICAL BEHAVIORS OF VCA IN COMPETITIVE LEARNING ENVIRONMENT

| Physical Behavior | Peedy Function |
|---|---|
| Persuade_Student_to_Think | Speak |

TABLE XIV.   VERBAL BEHAVIORS OF VCA IN THE COMPETITIVE LEARNING ENVIRONMENT

| Verbal Behavior | Peedy's Speaking |
|---|---|
| Persuade-Student-to-Think | I think you should take your time doing your assignments. |
| Persuade-Student-to-Think | No need to rush. Think them over! |

## VI.   EXPERIMENTAL RESULTS

For evaluating educational environment based on the proposed model, three educational environments were shown to thirty users and the users evaluated them. After that, they were asked to answer a questionnaire (Appendix A). According to users' responses, the following results are given (Figure 3-Figure 14):

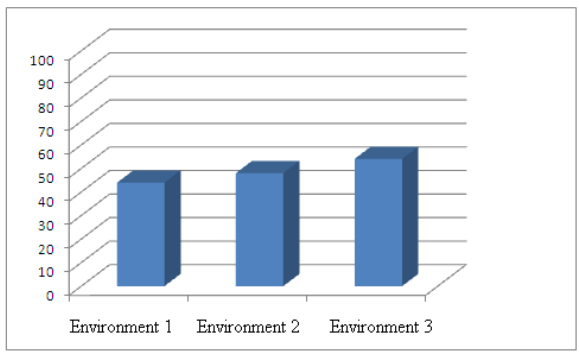

Figure 3.   Evaluating of Learning Rate

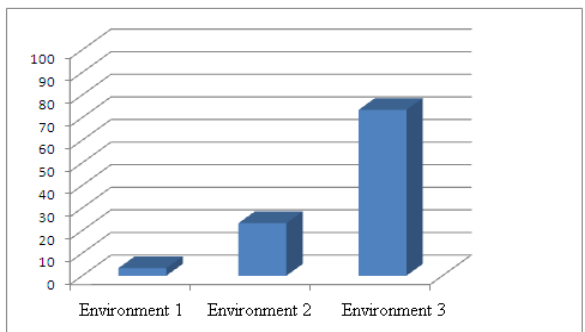

Figure 4.   Evaluating Attractiveness of Educational Environments

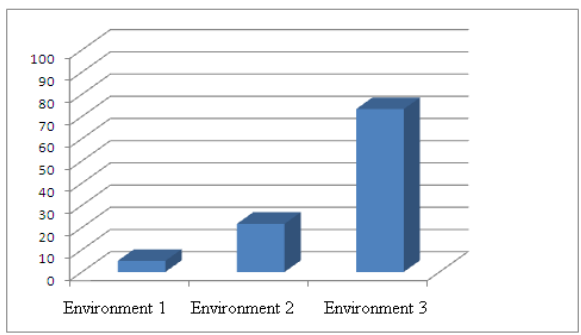

Figure 5.   Evaluating Interaction and User's Satisfaction of Educational Environments

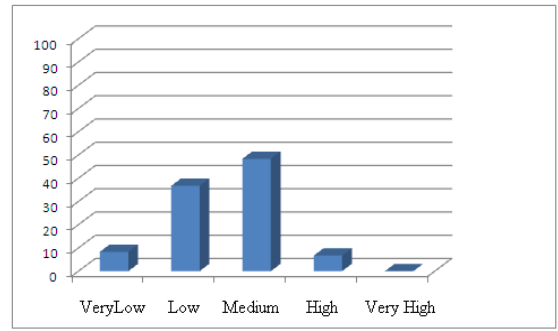

Figure 6.   Evaluating Interaction and User's Satisfaction Of Educational Environment

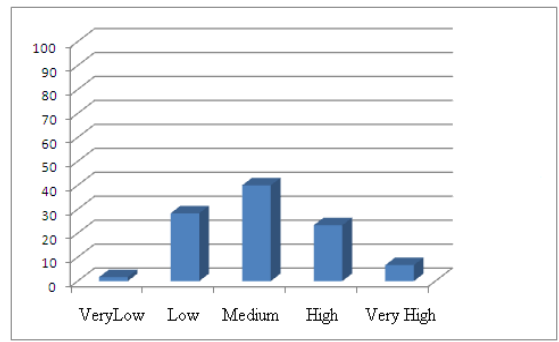

Figure 7.   Evaluating Interaction and User's Satisfaction of Educational Environment 2

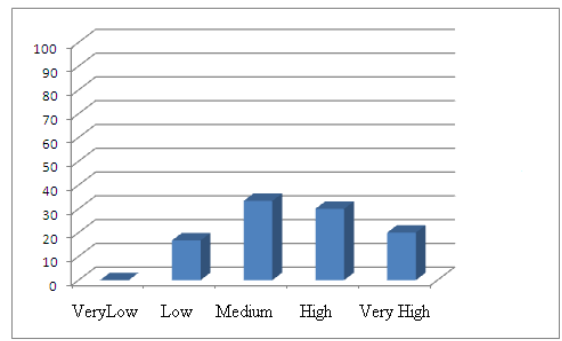

Figure 8.   Evaluating Interaction and User's Satisfaction of Educational Environment 3

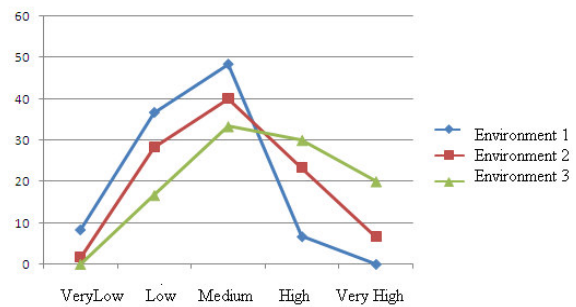

Figure 9.   Comparing User's Interaction and Satisfaction of Three Educational Environments.

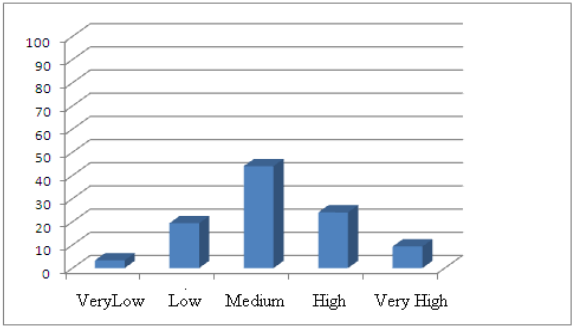

Figure 10. Evaluating VTA's Function in the Environment 2

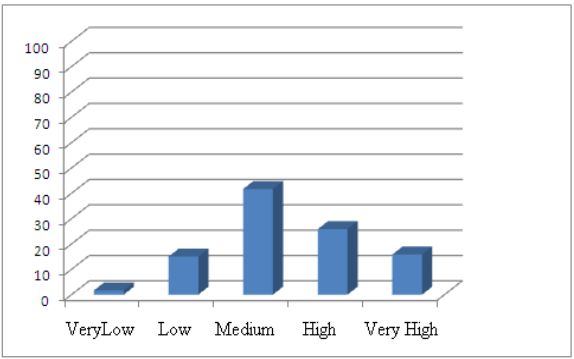

Figure 11. Evaluating VTA's Function in the Environment 3

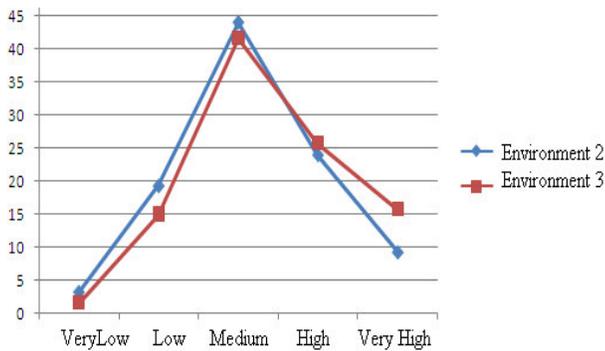

Figure 12. Comparing VTA's Function in the Environment 2 and 3

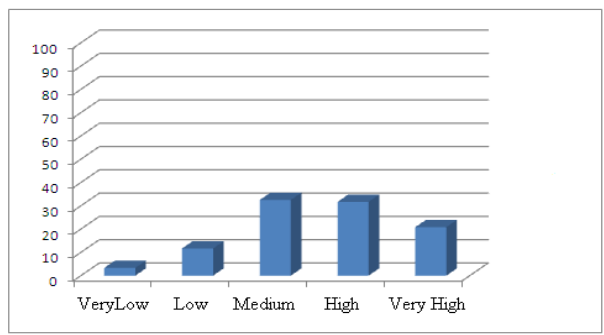

Figure 13. Evaluating VCA's Function

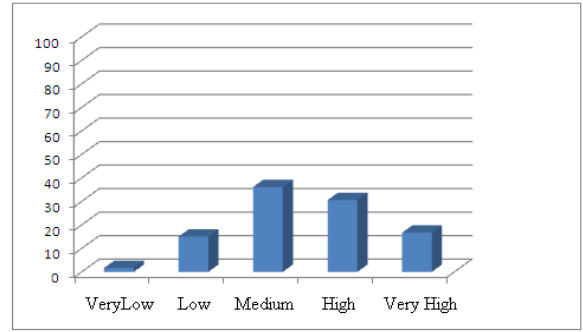

Figure 14. Evaluating the Effect of VCA's Presence in Learning Progress

## VII. CONCLUSION AND FUTURE WORKS

As the result of evaluation shows, users believe that the educational environment 3 is more attractive than the educational environments 1 and 2. Also, they believe the educational environment 2 is more attractive than the educational environment 1.

The result of the evaluation shows that the presence of the intelligent agents with features like human can increase the learning rate, and it has an important role in attracting them for the virtual educational environments. Comparing the educational environment 2 with the educational environment 3 showed that the presence of a VCA increases users' satisfaction and users' interaction with the environment.

Finally, the results show that VTA and VCA have considerable effect on improvement of learning process.

In the future, we will try to complete the system. We can do the following enhancements:

_Applying all dimensions of "MBTI":_ In the proposed model just two dimensions of MBTI were used for VCA simulation. For presenting this agent more realistically, two other dimensions can be added.

_Applying all emotions of the OCC model:_ In this model, just the emotions of the first and third branches of the OCC model were used. We can improve the model by using all of them.

_Applying the culture model:_ Nowadays, lots of researchers on the impacts of the culture on learning styles have been published. In the future, by adding this parameter to personality and emotion models, we can increase the environment credibility for the learners.

_Adding the learning feature to agents:_ In the implemented system, agents perceive the states according to the written rules of the system and do a suitable action for it. By adding the learning features to VTA and VCA and based on the learner's reactions in the environment, they can learn the rules.

_Completing the system:_ We can add some more functions to the system, such as lessons' classification according to their degree of hardness. In addition, it can increase the number of the VCA's in the learning environment.

# REFERENCES

[1] Abrahamian, E., Weinberg, J., Grady, M., and Michael Stanton, C. (2004). "The Effect of Personality-Aware Computer-Human Interfaces on Learning" , Journal of Universal Computer Science, Vol. 10, pp. 27-37.

[2] Al Masum, S.M. and Ishizuka, M. (2005). "An Affective Role Model of Software Agent for Effective Agent-Based e-learning by Interplaying between Emotions and Learning", WEBIST, USA, pp. 449-456.

[3] Anderson, T. and Elloumi, F. (2003). "Theory and practice of online learning".

[4] Berry, M. (2005). "A virtual learning environment in primary education".

[5] Botsios, S., Mitropoulou, V., Georgiou, D. and Panapakidis, I. (2006). "Design of Virtual co-Learner for Asynchronous Collaborative e-Learning". ICALT, pp. 406-407.

[6] Capretz L., F. (2002). "Implications of MBTI in Software Engineering Education", ACM SIGCSE Bulletin - inroads, ACM Press, New York, Vol. 34, pp. 134-137.

[7] Chaffar, S. and Frasson, C. (2004). "Using an Emotional Intelligent Agent to Improve the Learner's Performance", In Social and emotional intelligence in learning Environments workshop, 7th International Conference on Intelligent Tutoring System, Brazil, pp. 37-43.

[8] Chaffar, S., Cepeda, G. and Frasson, C. (2007). "Predicting the Learner's Emotional Reaction towards the Tutor's Intervention", 7th IEEE International Conference, Japan, pp. 639-641.

[9] Chalfoun, P., Chaffar, S. and Frasson, C. (2006). "Predicting the Emotional Reaction of the Learner with a Machine Learning Technique", Workshop on Motivaional and Affective Issues in ITS, International Conference on Intelligent Tutoring Systems, Taiwan

[10] Choi, K., Deek, F.P., and Im, I. (2008). "Exploring the Underlying Aspects of Pair Programming: The Impact of Personality", Journal of Information and Software Technology.

[11] Damasio, A.R. (1994). "Descartes' Error: Emotion, Reason, and the Human Brain ", Gosset/Putnam Press, New York.

[12] Dewar,T. and Whittington,D. (2000). "Online Learners and their Learning Strategies" , Journal of Educational Computing Research, Vol. 23, no. 4, pp. 415-433.

[13] Du, J., Zheng, Q., Li, H. and Yuan, W. (2005). "The Research of Mining Association Rules Between Personality and Behavior of Learner Under Web-Based Learning Environment", ICWL, pp. 406 – 417.

[14] Durling, D., Cross, N., and Johnson, J. (1996). "Personality and learning preferences of students in design and design-related disciplines". Proceedings of IDATER 96 (International Conference on Design and Technology Educational Research), Loughborough University, pp. 88-94.

[15] Ellis, S., and Whalen,    S. (1996). "Cooperative Learning: Getting Started", Scholastic.

[16] El-Nasr, M., Yen, J. and Ioerger, T. (2000), "FLAME – fuzzy logic adaptive model of emotions", Autonomous Agents and Multi-Agent Systems, Vol. 3, no.3, pp. 219-57.

[17] Fatahi, S., Ghasem-Aghaee, N., and Kazemifard, M. (2008). "Design an Expert System for Virtual Classmate Agent (VCA)". Proceedings of World Congress Engineering 2008, U.K, London.

[18] Fatahi, S., M. Kazemifard, and N. Ghasem-Aghaee. 2009. "Design and Implementation of an E-Learning Model by Considering Learner's Personality and Emotions". In Advances in Electrical Engineering and Computational Science, Netherlands: Springer, Vol. 39, pp. 423-434.

[19] Ghasem-Aghaee, N., Fatahi, S., and T.I. Ören. (2008). "Agents with Personality and Emotional Filters for an E-learning Environment", Proceedings of Spring Agent Directed Simulation Conference, Ottawa, Canada.

[20] Harati Zadeh, S., bagheri Shouraki, S. and Halavati, R. (2006). "Emotional Behavior: A Resource Management Approach", Journal of SAGE, Vol. 14, pp. 357-380.

[21] Hartmann, P. (2006). "The Five-Factor Model: Psychometric, biological and practical perspectives".

[22] Higgs, M. (2001). "Is there a relationship between the Myers-Briggs type indicator and emotional intelligence?", Journal of Managerial Psychology, Vol. 16, no. 7, pp. 509-533.

[23] http://www.murraystate.edu

[24] Jessee, S.A., ONeill, P.N. and Dosch, R.O. (2006). "Matching Student Personality Types and Learning Preferences to Teaching Methodologies", Journal of Dental Education, Vol. 70, pp. 644-651.

[25] Ju, W., Nickell, S., Eng, K. and Nass, C. (2005). "Influence of colearner agent behavior on learner performance and attitudes", CHI '05 extended abstracts on Human factors in computing systems, ACM Press, Portland, OR, USA.

[26] Kazemifard, M., Ghasem-Aghaee, N., and Oren, T. I. (2006). "An Event-based Implementation of Emotional Agents". In Proceeding of the Summer Simulation Conference (SCSC'06), Calgary, Canada,pp. 63-67.

[27] Kazemifard, M., and Ghasem-Aghaee, N. (2007). "Virtual Tutor". In Proceeding of the 15th Iranian Conference on Electrical Engineering (ICEE'07), Tehran, Iran,pp. 358-363.

[28] Kestern, A.J.(2001)."A supervised machine-learning approach to artificial emotions". Master's thesis, Department of Computer Sience, University of Twent.

[29] Kort, B. and Reilly, R. (2001). "Analytical Models of Emotions, Learning and Relationships: Towards an Affect-sensitive Cognitive Machine". MIT Media Lab Tech Report, no. 548.

[30] Kshirsagar, S. and Magnenat-Thalmann, N. (2002). "A multilayer personality model", Proceedings of the 2nd international symposium on Smart graphics, Hawthorne, New York pp.107-115.

[31] Lang, H.G., Stinson, M.S., Kavanagh, F., Liu, Y., and Basile, M. (1999). "Learning styles of deaf college students and instructors' teaching emphases". Journal of Deaf Studies and Deaf Education, Vol. 4, pp. 16-27.

[32] Li, Y.S., Chen, P.S., and Tsai, S.J. (2007). "A comparison of the learning styles among different nursing programs in Taiwan: Implications for nursing education", Journal of Nurse Education Today. Vol. 28, no. 1, pp. 70-76.

[33] Logan, K., and Thomas, P. (2002). "Learning Styles in Distance Education Students Learning to Program", Proceedings of 14th Workshop of the Psychology of Programming Interest Group, Brunel University, pp. 29-44.

[34] Maria, K.A. and Zitar, R.A. (2007). "Emotional Agents: A Model and an Application", Journal of Information and Software Technology, Elsevier Publishing, Vol. 49, no. 6, pp. 695–716.

[35] Marin , B.F., Hunger, A. and Werner, S. (2006). "Corroborating Emotion Theory with Role Theory and Agent Technology: a Framework for Designing Emotional Agents as Tutoring Entities", Journal of Networks, Vol. 1, pp. 29-40.

[36] Miranda, J.M and Aldea, A. (2005). "Emotions in Human and Artificial Intelligence", Journal of Computers in Human Behaviour, Vol. 21, no. 2, pp. 323-341.

[37] Morishima, Y., Nakajima, H., Brave, S., Yamada, R., Maldonado, H., Nass, C. and Kawaji, S. (2004), "The Role of Affect and Sociality in the Agent-Based Collaborative Learning System", In Tutorial and Research Workshop, New York, pp. 265-275.

[38] Ortony, A., Clore, G. L. and Collins, A .1988. "The Cognitive Structure of Emotions", Cambridge University Press, Cambridge, UK.

[39] Peslak, A.R. (2006). "The impact of personality on information technology team projects", Proceedings of the 2006 ACM SIGMIS CPR conference on computer personnel research: Forty four years of computer personnel research: achievements, challenges & the future, Claremont, California, USA.

[40] Rushton, S., Morgan, J. and Richard, M. (2007). "Teacher's Myers-Briggspersonality profiles: Identifying effective teacher personality traits", Journal of Teaching and Teacher Education, Vol. 23, pp. 432-441.

[41] Sarmento, L.M. (2004). "An Emotion-Based Agent Architecture", Master Thesis in Artificial Intelligence and Computing, Universidade do Porto.

[42] Schultz, D. P., and Schultz, S.E., (2008) "Theories of Personality", edition 4.

[43] Shermis, M.D. and Lombard, D. (1998). "Effects of computer-based test administration on test anxiety and performance". Journal of Computers in Human Behavior, Vol. 14, pp. 111-123.

[44] Vinayagamoorthy, V., Gillies , M., Steed, A., Tanguy, E., Pan, X., Loscos, C., and Slater,M. (2006). Building Expression into Virtual Characters" , In Eurographics 2006. Vienna: Proceeding of Eurographics.

[45] Vincent, A. and Ross, D. (2001). "Personalize training: determine learning styles, personality types and multiple intelligences online", The Learning Organization, Vol. 8, no. 1, pp. 36±43 ISSN 0969-6474.

[46] Xiangjie, Q., Zhiliang, W., Jun, Y. and Xiuyan, M. (2006). "An Affective Intelligent Tutoring System Based on Artificial Psychology". Beijing University of Science and Technology, China, pp. 402-405.

[47] Yeung A., Read J. and Schmid S. (2005). "Students' learning styles and academic performance in first year chemistry". Proceedings of the Blended Learning in Science Teaching and Learning Symposium. Sydney,NSW: UniServe Science, pp. 137–42.

Appendix A.

A few sample questions from the questionnaire for the user's satisfaction of the system.

(i) Which of the educational environments is more attractive?
   (a) The educational environment 1(simple educational environment)
   (b) The educational environment 2(educational environment with emotional virtual tutor)
   (c) The educational environment 3(educational environment with virtual tutor and classmate agent who has emotions and personality)

(ii) Your satisfaction level of the educational environment 3(educational environment with a virtual tutor and a classmate agent with emotions and personality)
   a) Very Low       b) Low       c) Medium       d) High
   e) Very High

(iii) Your satisfaction level of the virtual tutor's helps in the educational environment 2:
   a) Very Low       b) Low       c) Medium       d) High
   e) Very High

(iv) Your satisfaction level of the virtual classmate's encouragements in the educational environment 2:
   a) Very Low       b) Low       c) Medium       d) High
   e) Very High

(v) Your satisfaction level of the virtual classmate's encouragements in the educational environment 3:
   a) Very Low       b) Low       c) Medium       d) High
   e) Very High

(vi) How do you evaluate the presence of the virtual classmate agent?
   a) Poor            b) Fair       c) Acceptable       d) Good
   e) Excellent

(vii) Your evaluation of the effect of the cooperation with the virtual classmate agent in learning progress:
   a) Very Low       b) Low       c) Medium       d) High
   e) Very High

(viii) Your evaluation of the effect of the competition  with the virtual classmate agent in learning progress:
   a) Very Low       b) Low       c) Medium       d) High
   e) Very High

**Somayeh Fatahi** received her B.Sc. and M.Sc. degrees in Computer Engineering from Razi University, and Isfahan University, Iran, in 2006 and 2008 respectively. She is Lecturer in the Department of Computer Engineering at Kermanshah University of Technology. She also teaches in Razi University, University of Applied Science and Technology, Payam Noor University of Kermanshah, Islamic Azad University of Kermanshah, Institute of Higher Education of Kermanshah Jahad-Daneshgahi.
Her research activities include (1) simulation of agents with dynamic personality and emotions (2) Computational cognitive modeling (3) Simulation and formalization of cognitive processes (4) Multi Agent Systems (5) Modeling of Human Behavior  (6) Fuzzy Expert Systems.

**Dr. Nasser Ghasem-Aghaee** is a co-founder of Sheikhbahaee University of Higher Education in Isfahan, Iran, as well as Professor in the Department of Computer Engineering at both the Isfahan University and Sheikhbahaee University. In 1993-1994 and 2002- 2003, he has been visiting Professor at the Ottawa Center of the McLeod Institute for Simulation Sciences at the School of Information Technology and Engineering at the University of Ottawa. He has been active in simulation since 1984. His research interests are modeling and simulation, cognitive simulation (including simulation of human behaviour by fuzzy agents, agents with dynamic personality and emotions, artificial intelligence, expert systems, fuzzy logic, object-oriented analysis and design, multi-agent systems and their applications. He published more than 100 documents in Journals and Conferences.